\documentclass[%
superscriptaddress,
twocolumn,
10pt,
aps,
prb,
floatfix,
]{revtex4-1}

\usepackage[version=3]{mhchem} 
\usepackage[T1]{fontenc}       
\usepackage{graphicx}
\usepackage{dcolumn}
\usepackage{bm}
\usepackage{epsfig}
\usepackage{natbib}
\usepackage{amsmath}
\usepackage{mathrsfs}
\usepackage{amssymb}
\usepackage{times}
\usepackage{psfrag}
\usepackage[english]{babel}
\usepackage[]{subfigure}
\usepackage{multirow}
\usepackage{color}
\usepackage{multirow}
\usepackage{xcolor,soul}

\bibpunct{[}{]}{,}{n}{}{}   

\newcommand{\comm}[1]{}

\begin{document}
	
\title{Proton-transfer induced fluorescence in self assembled short peptides.}

\author{Sijo K. Joseph}
\affiliation{Department of Physical Electronics, Tel-Aviv University, Tel-Aviv 69978, Israel.}
\affiliation{The Sackler Center for Computational Molecular and Materials Science, Tel-Aviv University, Tel-Aviv 69978, Israel.}
\author{Natalia Kuritz}
\affiliation{Department of Physical Electronics, Tel-Aviv University, Tel-Aviv 69978, Israel.}
\author{Eldad Yahel}
\affiliation{Department of Physical Electronics, Tel-Aviv University, Tel-Aviv 69978, Israel.}
\author{Nadezda Lapshina}
\affiliation{Department of Physical Electronics, Tel-Aviv University, Tel-Aviv 69978, Israel.}
\author{Gil Rosenman}
\affiliation{Department of Physical Electronics, Tel-Aviv University, Tel-Aviv 69978, Israel.}
\author{Amir Natan}
\email{amirnatan@post.tau.ac.il }
\affiliation{Department of Physical Electronics, Tel-Aviv University, Tel-Aviv 69978, Israel.}
\affiliation{The Sackler Center for Computational Molecular and Materials Science, Tel-Aviv University, Tel-Aviv 69978, Israel.}

\begin{abstract}
We employ Molecular Dynamics (MD) and Time-Dependent Density Functional Theory (TDDFT)
to explore the fluorescence of Hydrogen-bonded dimer and trimer structures of Cyclic FF (Phe-
Phe) molecules. We show that in some of these configurations a photon can induce either an \textit{intra}-molecular proton transfer, or an \textit{inter}-molecular proton transfer that can occur in the excited S1 and S2 states. This proton transfer, taking place within the Hydrogen bond, leads to a significant red-shift that can explain the experimentally observed visible fluorescence in biological and bioinspired peptide nanostructures with a $\beta$-sheet biomolecular arrangement. Finally, we also show that such proton transfer is highly sensitive to the geometrical bonding of the dimers and trimers, and that it occurs only in specific configurations allowed by the formation of Hydrogen bonds.   
\end{abstract}

\maketitle

\section{Introduction}
The discovery of the Green Fluorescent Protein (GFP) in jellyfish bio-luminescence by Osamu
Shimomura has opened the door to new technologies of bio-compatible markers that can be used in-vitro and in-vivo~\cite{tsien1998green,Wolfbeis_Book,Chalfie802}. In general, both GFP and fluorescence have important technological applications, such as monitoring gene expression and protein localization in living organisms, as well as detection of migration and invasion of cancer cells in vivo~\cite{Book_RM_Hoffman}. Fluorescence can also shed light on the structure of some important biological systems, such as amyloid fibrils, that play a critical role in the Alzheimer disease mechanism \cite{Alzheimer_Bioquimica,Dobson2001,HamleyAmloid2007,
Alzheimer_Cell_13,delMercato2007,chan2013protein,grisanti2017computational,
sharpe2011solid,Xue2017}, or natural spiders' silk~\cite{AssemblySpiderSilk1,SpiderBioPhysJ,SpiderNanoConfin,KetenPRL}, 
as well as within nanotechnology applications such as the field of peptide nanophotonics~\cite{AHandelMan2018}.

In recent years there have been many observations of an emergent fluorescence phenomenon in self-
assembled short peptide aggregates ~\cite{Aggr_Induce_FL2017,Aggr_FL2009,AggrInduce1,AggrInduce2,AggrInduce3}, and in peptide and protein fibers, including the important case of amyloid fibrils~\cite{delMercato2007,chan2013protein,grisanti2017computational,sharpe2011solid,Xue2017}. It has been observed that the excitation of such molecular structures in the UV spectrum can yield fluorescence in the visible range, i.e. the emission of red, green 
and blue colors.

It is generally agreed that the secondary structure, and, specifically, the formation of structures such as $\beta$-sheets that are rich with Hydrogen bonds, play a critical role in the appearance of visible range fluorescence~\cite{HydrogenBondFluro2012,pinotsi2016proton,flur_ptran_htrans2017}. In particular, this phenomenon is less widely observed 
or does not exist in the monomers or in other forms of secondary structures that are based on the same monomers~\cite{handelman2016reconstructive}. 

In 2004 Shukla et al.~\cite{shukla2004novel} suggested that Hydrogen bonds may
led new delocalized electronic states that might be related to the
appearance of fluorescence. Pinotsi et al.~\cite{pinotsi2013label} and Chan et al.~\cite{chan2013protein} have shown that fluorescence is related to protein aggregation in amyloid fibrils, and to the formation of $\beta$-sheet regions that are rich with Hydrogen bonds. 
Recently, fluorescence in the visible range was demonstrated for various types of self-assembled fibers of short peptides structures, that are also supported with a $\beta$-sheet like organization~\cite{handelman2016reconstructive, Nadav_Gill09,Amdursky_2011,chan2013protein,BookSelfassembly11,GanFF_PL2017,Aggr_Induce_FL2017,
delMercato2007,grisanti2017computational,sharpe2011solid,Xue2017,apter2018peptide}.

Excited state \textit{intra}-molecular proton transfer (ESIPT) is a known mechanism that can lead to fluorescence in many molecules 
~\cite{RefESIPT_1,RefESIPT_2,RefESIPT_3,RefESIPT_4,jagadesan2017excited}. 
Proton transfer between different amino acids is also considered to be the leading mechanism in GFP~\cite{tsien1998green,fang2009mapping}. 
Pinotsi et al.~\cite{pinotsi2016proton} and 
Grisanti et al.~\cite{grisanti2017computational} have shown, using Molecular Dynamics (MD) and Time-Dependent Density Functional Theory 
(TDDFT), that an \textit{inter}-molecular proton transfer can happen along the Hydrogen bond between the N terminal and the C terminal of model amyloid fibril proteins in a crystalline structure.
They have also shown that this proton transfer affects the ground and excited state properties and, hence, could be the main factor behind the fluorescence in such structures.  

In this work, we explore cyclic Phe-Phe (Cyclic-FF) peptide dimers and trimers in the ground and excited state geometries, in order to evaluate the absorption and emission spectra, respectively. 
We use MD and TDDFT~\cite{runge1984density} calculations for this 
analysis. We show that in some geometries of dimers and trimers, {\it intra}-molecular and  \textit{inter}-molecular proton transfer through a Hydrogen bond can happen in the excited state. We show that this leads to a dramatic change in the optical properties of the dimers and trimers, leading to a large red-shift of the electronic gap. This shift can explain the appearance of fluorescence in the visible range. While the calculations are focused on Cyclic-FF, the mechanism of proton transfer is not directly related to the aromatic side groups and is probably of a general nature. Therefore, we expect to find similar behavior in aggregates of other peptides.  

\section{Methods and computational protocol}
We used TDDFT~\cite{CA_Ullrich_Book,runge1984density} calculations to investigate the excited state electronic properties of Cyclic-FF monomers, dimers, and trimers. Linear response theory with the Tamm-Dancoff Approximation (TDA) was used to evaluate the excited state energies and oscillator strengths ($|S(\lambda)|^2$) 
~\cite{CasidaChapter,CasidaReview,TDDFT_Tamm_Dancoff}. We used the 6-311G** basis-set with the B3LYP functional and Grimme dispersion 
Van der Waals corrections (B3LYP-D3)~\cite{GrimmeDFTD3}. The excited state minimum energy geometry was determined from energy Hessian calculations, as implemented in the Q-Chem code~\cite{QChem4Ref}. The protocol for finding candidate dimer and trimer structures was the following:

\begin{itemize}
	\item Several initial configurations for dimers and trimers were produced, either manually or from MD simulations.
	\item Each configuration was further geometrically relaxed to its ground state energy minimum using the B3LYP-D3 functional and the 6-311G** basis set.
	\item The absorption spectrum was calculated with linear response TDDFT.
	\item The minimum energy geometries for the first and second excited state (S1 and S2) were calculated, using the B3LYP-D3 functional and the 6-311G** basis set.
	\item The emission spectrum was calculated with linear response TDDFT for the excited states minimum energy geometries. 
\end{itemize}

The absorption and emission spectra were determined from the computed oscillator strength. We determined the first 40 excited electronic states for all the geometries in this study. The oscillator strengths were convoluted with a Lorentzian curve to obtain the desired continuous spectra.

The MD simulations were performed using QMDFF, which is a general quantum mechanically derived
force field \cite{Grimme2014a} based on ORCA \cite{Neese2012} DFT calculations. For the ORCA DFT calculations we employed the B3LYP functional with the Grimme dispersion correction (B3LYP-D3) and the TZVPP basis set. 
Based on these calculations, the QMDFF program was used to produce a force field. The molecular dynamics simulations were performed within the framework of QMSIM, which is a molecular dynamics simulation program of nonperiodic and periodic systems developed by S. Grimme. The 
MD procedure contained a few (2-4) molecules in periodic boundary conditions, which were optimized, and then, simulated the dynamics of 
typically 3-5 cycles of 1000 time steps at a high temperature (480K) and 1000 time steps at room temperature (300K), where each time step was 0.5-1ps in duration.

\section{Results and discussion}
In this section we present the results for Cyclic-FF dimers and trimers.

\subsection{Cyclic-FF Dimer }
We used the MD procedure as well as a manual guess structure, for the Hydrogen-bonded dimers' initial
configurations, which were further relaxed with DFT. The energy dependence of the MD procedure of the Hydrogen-bonded dimers' initial configurations, at each time step, is shown in Fig.1. We also show the two low energy configurations that were eventually selected for the TDDFT analysis.
\begin{figure}[!htbp]
\centering
{\includegraphics[width=1.0\linewidth]{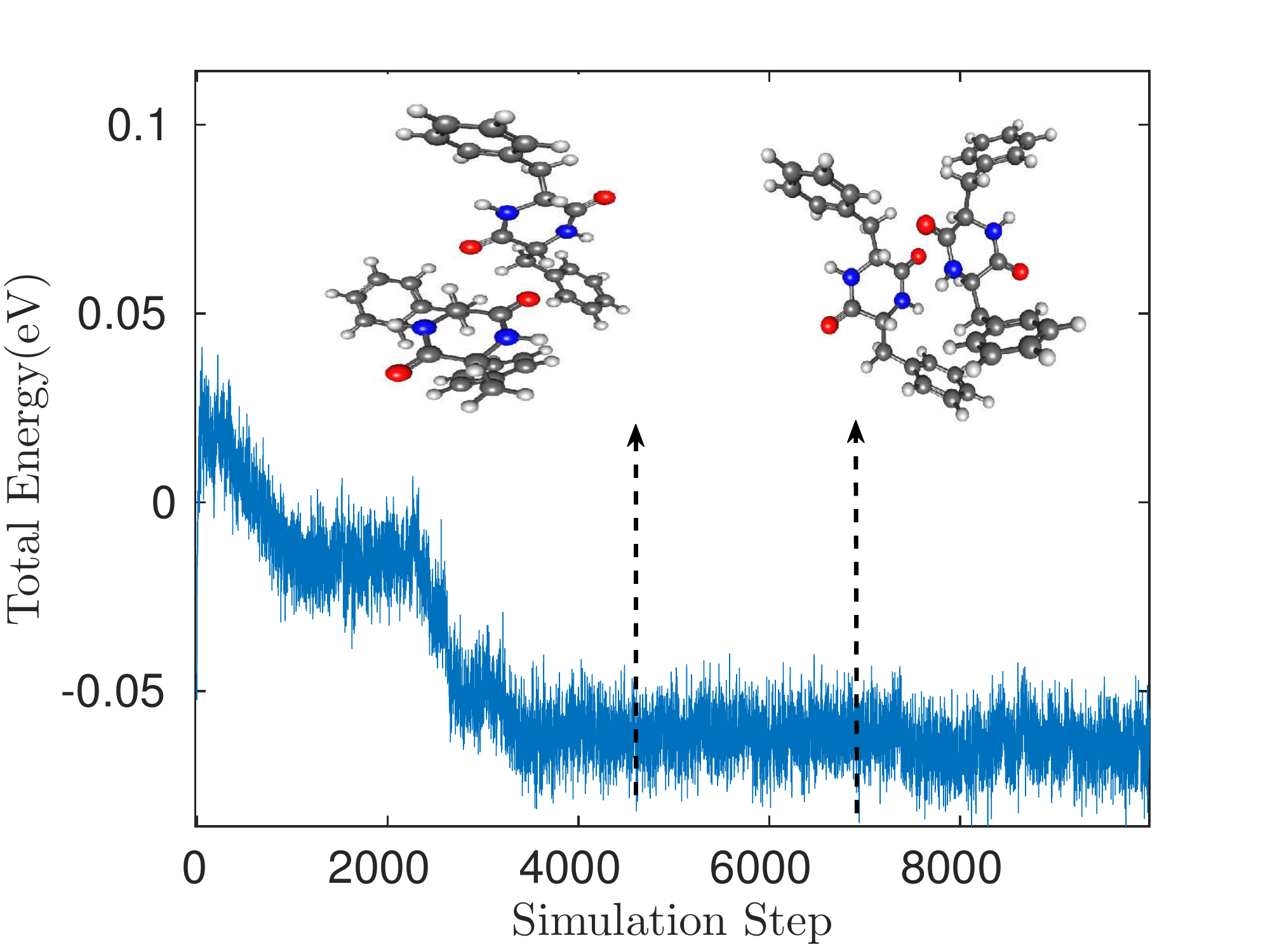}}
\caption{Energy as a function of the MD simulation step for the Hydrogen-bonded dimers. Two of the lowest energy structures, which were selected for further analysis, are marked by the black vertical arrows, and their configurations are shown with a ball and stick atomistic model.}
\label{fig:dimer_MD}
\end{figure}

\begin{figure*}[!htbp] 
\includegraphics[height=5.0cm,width=16cm]{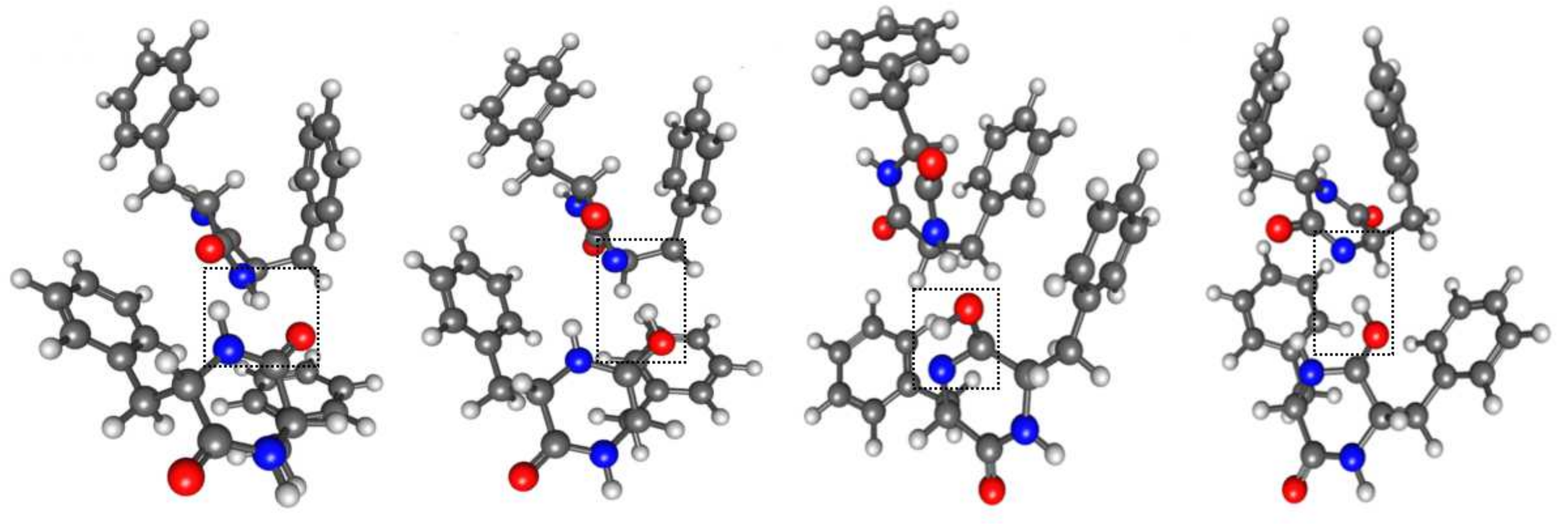}
\caption{The ground state geometry, similarly the first and second excited state geometries of the Cyclic-FF Dimer, are shown. Here, (a) shows the ground state geometry, (b) and (c) show the inter-molecular proton transfer and intra-molecular proton transfer from the ground state to the first excited state $S_{1}$, respectively, whereas (d) refers to the inter-molecular proton transfer from the ground state to the second excited state $S_2$ . In all three cases, the ground state C=O bond is transformed into a C-O-H bond via an intra- or inter-molecular proton transfer from a nearby N-H bond. The dashed line rectangle shows the location of the Hydrogen bond.}
\label{CFFDimfig2}
\end{figure*}

\begin{table*}[!htbp]
	
	\begin{tabular}{ | l | l | l |} \hline
		Cyclic-FF Dimer & Cyclic-FF Dimer Ground State   & Binding Energy  \\ 
		Ground State Geometry &  Energy in Hartree ($Ha$)& per Molecule in Hartree ($Ha$)\\ \hline
		MD Geometry 4606   & -1914.097  & -0.0201 \\ \hline
		MD Geometry 6941   & -1914.096  & -0.0197 \\ \hline
		Manual guess  & -1914.108  & -0.0259 \\ \hline
	\end{tabular}
	\caption{Binding energies for different dimer geometries}
	\label{table:dimer_binding_energy}
\end{table*}

\begin{table*}[!htbp]
\begin{tabular}{ | l | l | l |l|} \hline
Estimated Dimer & Ground State &  Excited State & Energy Gap in $eV$ \\
Geometry & Energy in $H_{a}$  & Energy in $H_{a}$ & ($\Delta E_{n}=E_{n}-E_{0}$)\\ \hline
                              &                     &                    &                             \\
Ground State Optimized        &  $E_{0}$= $-1914.1082$& $E_{1}=-1913.9135$ & $\Delta E_{1}= 5.297 \,$  \\ 
Geometry                      &                     & $E_{2}=-1913.9132$ & $\Delta E_{2}= 5.307 \,$  \\ \hline
First Excited State           &                     &                    &                             \\
Optimization (Inter-Molecular & $E_{0}$= $-1913.9854$ & $E_{1}=-1913.9790$ & $\Delta E_{1}= 0.174 \,$  \\
Proton Transfer) (Geometry-I)   &                     & $E_{2}=-1913.9642$ & $\Delta E_{2}= 0.576 \,$  \\\hline
First Excited State           &                     &                    &                             \\
Optimization (Intra-Molecular & $E_{0}$= $-1914.0421$ & $E_{1}=-1913.9464$ & $\Delta E_{1}= 2.604 \,$  \\ 
Proton Transfer) (Geometry-II)  &                     & $E_{2}=-1913.9077$ & $\Delta E_{2}= 3.657 \,$  \\ \hline
Second Excited State          &                     &                    &                             \\ 
Optimization (Inter-Molecular & $E_{0}$= $-1913.9846$ & $E_{1}=-1913.973$  &  $\Delta E_{1}= 0.327 \,$ \\          
Proton Transfer)              &                     & $E_{2}=-1913.972$  &  $\Delta E_{2}= 0.334 \,$ \\ \hline
\end{tabular}
\caption{Proton-transferred dimer ground and excited state energy gap}
\label{table:dimer_excitation_energies}
\end{table*}

The results of the ground state geometry optimization are shown in Table 1. The calculated Cyclic-FF Monomer ground state energy was $-957.028$ Hartree (Ha).

According to Table 1, the most stable configuration that we obtained was based on the initial estimation, with a binding energy of 0.0259 Hartree (0.7048eV), or 16.2525 Kcal/mol, per monomer. We observed two Hydrogen bonds in the optimized dimer configuration. Hence, the binding energy for a single Hydrogen bond is 8.1263 Kcal/mol. This value is close to the reported value of $5-8$ Kcal/mol for a single Hydrogen bond in $\beta$-sheet structures ~\cite{sheu2003energetics}. The MD
generated structures are slightly higher in energy and have lower dimer binding energies, with the difference being around $2K_{B}T$ at room temperature.

We first analyzed the behavior of the manually generated structure under light excitation. It is evident from Table 1 that the first and second excited states ($S_1$ and $S_2$) are degenerate. We performed the following minimization processes for the geometry of the excited state: for the $S_1$ first excited state we performed one minimization process by first operating with the smaller 3-21G basis set and then continuing with the 6-311G** basis. This two-step procedure  led to an {\it inter}-molecular 
proton transfer state with a gap of 0.17eV; we call this "Geometry-I" and describe it in Fig.~\ref{CFFDimfig2}b. 
In addition, we performed another minimization 
process for the $S_1$ state, where we started from the ground state geometry and did the excited state geometry minimization with 
the 6-311G** basis from the start. This led to another energy minimum, which we call "Geometry-II" for further reference ; this geometry is shown in Fig.~\ref{CFFDimfig2}c. In this case, we got an {\it intra}- 
molecular proton transfer state, leading to a gap of 2.6eV. Geometry-I is 0.8eV lower in excited state energy ($E_1$) in comparison to 
Geometry-II. Here, there could be some barriers that did not let the system relax into Geometry-I when operating on the larger basis initially. However, if we also consider proton tunneling~\cite{ProtonTunnel1,ProtonTunnel2,ProtonTunnel3} we can assume that both states can happen.

In addition to minimization of geometry in the $S_1$ state, 
we performed minimization of geometry in the $S_2$ state; here, we started 
from the ground state geometry and performed 
the minimization with the 6-311G** basis. 
The resulting minimum is again an {\it inter}- molecular proton transfer state - similar to Geometry-I of the $S_1$ minimization. The observed gap in this minimum was found to be 0.33eV.

In Fig. 2(a), it can
be observed that a Hydrogen atom is attached to the Nitrogen atom of the bottom molecule (see the
small box around the Hydrogen), while in the Geometry-I excited state (Fig. 2c), this Hydrogen atom is
transferred to the nearby Oxygen atom of the second molecule, resulting in the formation of a C-O-H bond.
 This {\it inter}-proton transfer can also be interpreted as a
transition of the dimer from the Lactam form to the Lactim form ~\cite{TautomerBook,BookTautomerism,RCTautomerism,lactam_lactim_experi}. Geometry-II (Fig2b) also shows a proton transfer, but now it is an {\it intra}-proton transfer.
The change in the Hydrogen atom location leads to a high reduction of 
the de-excitation energy, which results in a significant red-shift in the emitted wavelength. A similar proton transfer mechanism is also demonstrated for the $S_{2}$ excited state
minimization, as shown in Fig.~2d.

The proton transfer is illustrated further in Fig.~3a, where the distances between the Hydrogen atom and the neighbor Oxygen and Nitrogen atoms are shown as a function of the S$_2$ energy minimization step. Fig.~3b shows the ground and excited states' energies as a function of the minimization step; while there are some jumps in the energy, it shows a large decrease in the S$_2$ and S$_1$ excited states in parallel to the proton transfer. At the same time, the ground state energy, S$_0$, increases, leading to a very small gap. The minimization becomes noisy after the proton transfer as the S$_2$, S$_1$ and S$_0$ states become almost degenerate, which makes the Born-Oppenheimer approximation less justified. We discuss this further in the supporting information (SI).

\begin{figure}[!htbp]
\centering
{\includegraphics[width=1.0\linewidth,trim={1cm 3cm 0 2cm},clip]{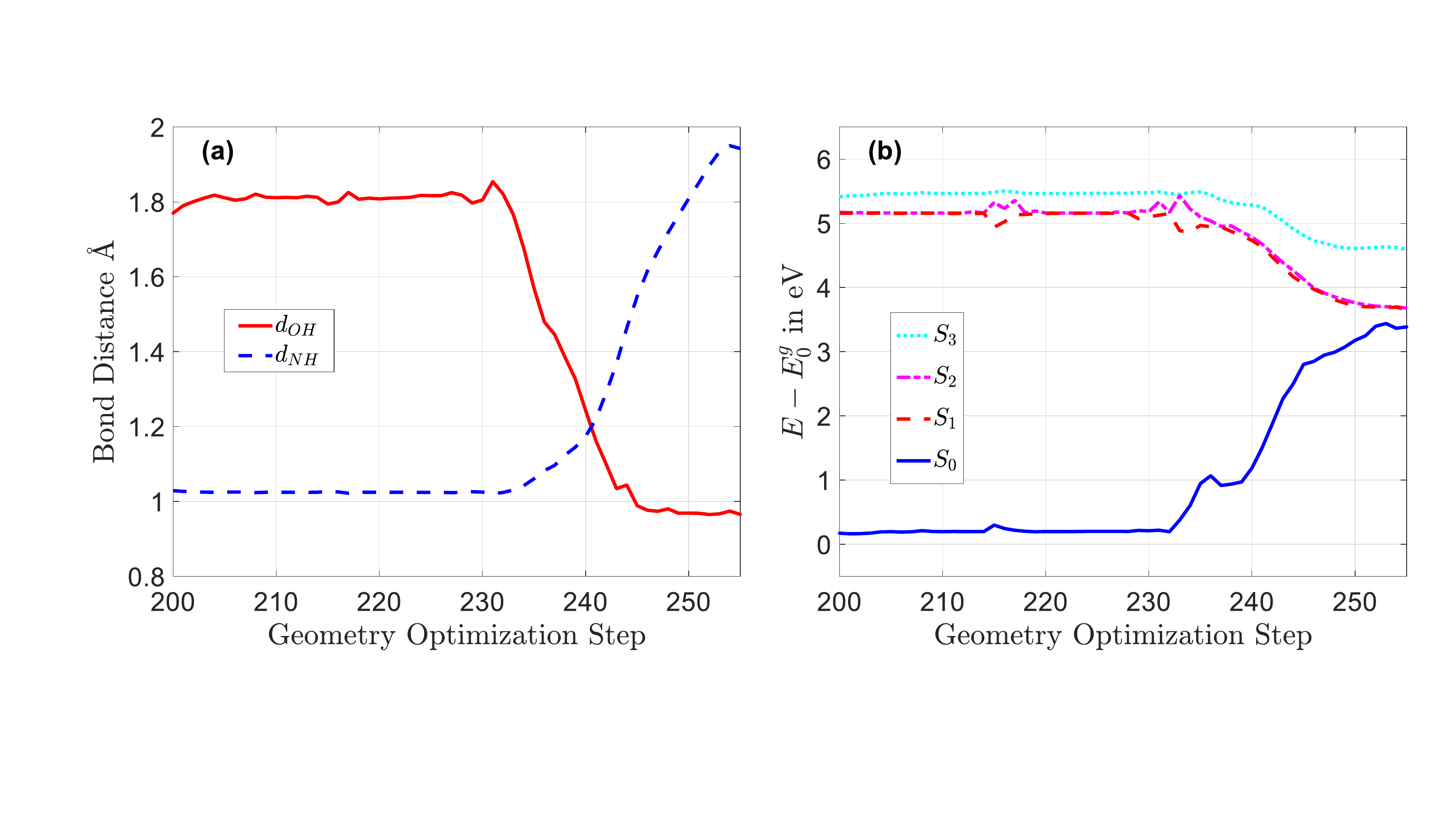}}
	\caption{S$_2$ minimization: (a) OH and NH distances as a function of minimization step; (b) ground and a few excited state energies as a function of minimization step. The full minimization graph is shown in the SI.}
	\label{fig:fig3}
\end{figure}

The fluorescence mechanism can be
interpreted as follows. First, the dimer absorbs a UV-light photon, transitioning to a higher electronic excited state, but still without nuclear motion, hence staying in the ground state geometry. Then, the molecule's nuclei relax their 
energy via a proton-transfer mechanism in the excited electronic state. 
It can be seen that the proton transfer leads to a significant change in both the S$_0$ and S$_n$ electronic states, and that the gap
is significantly reduced. Once the proton-transferred geometry is reached, the gap between the
excited state and the ground state is significantly lower, i.e. possibly in the visible, or even the IR, range. Hence, the dimer makes a transition from the proton-transferred excited state to the ground state (i.e. $S_{n} \to S_{0}$), yielding
fluorescence at a much longer wavelength. The resulting calculated band gap, in both the $S_{1}\to S_{0}$
and $S_{2} \to S_{0}$ inter-molecular proton transfer states, is in the IR range and not in the visible range. However, there could be several possible ways to get a gap in the visible range. 
First, the TDDFT calculation might have under-estimated the gap. Second, and perhaps more importantly, there could be two additional mechanisms that are possible; (a) a similar transition of a higher state 
$S_{n} \to S_{0}$ that also has a proton transfer, such transitions can also explain excitation dependent fluorescence, where the fluorescence wavelength depends on the excitation wavelength (b) a delayed fluorescence mechanism ~\cite{delayedfluro1}, i.e. the electron is first excited to $S_{1}$ , but might climb to higher electronic states during the geometrical relaxation. Those mechanisms certainly require additional research, both experimentally and theoretically. 
%

\begin{figure}[!htbp]
\includegraphics[width=1.0\linewidth]{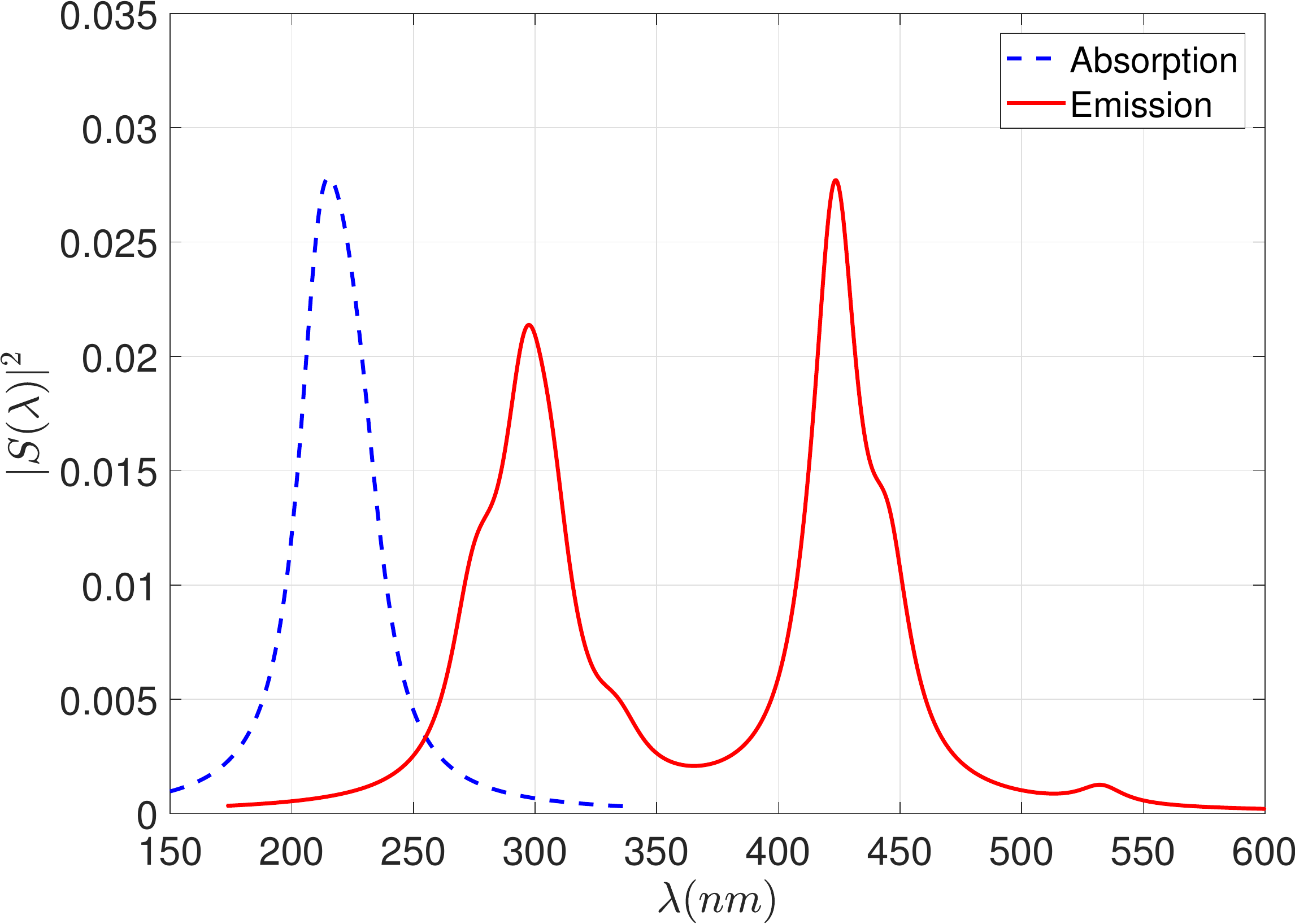}
\caption{The absorption (dashed blue) and emission (solid red) spectra of the Cyclic-FF Molecule, computed with B3LYP-D3/6-311G**. The solid curve shows the emission spectra in a proton-
transferred excited state geometry; a significant shift to the visible range is obtained compared to absorption spectra. Here $|S(\lambda)|^2$ is the dimensionless oscillator strength.}
\label{fig:manual_dimer_spectra}
\end{figure}
\begin{figure}[!htbp]
	\centering
	\includegraphics[width=1.0\linewidth,trim={1cm 3.5cm 0 2.5cm},clip]{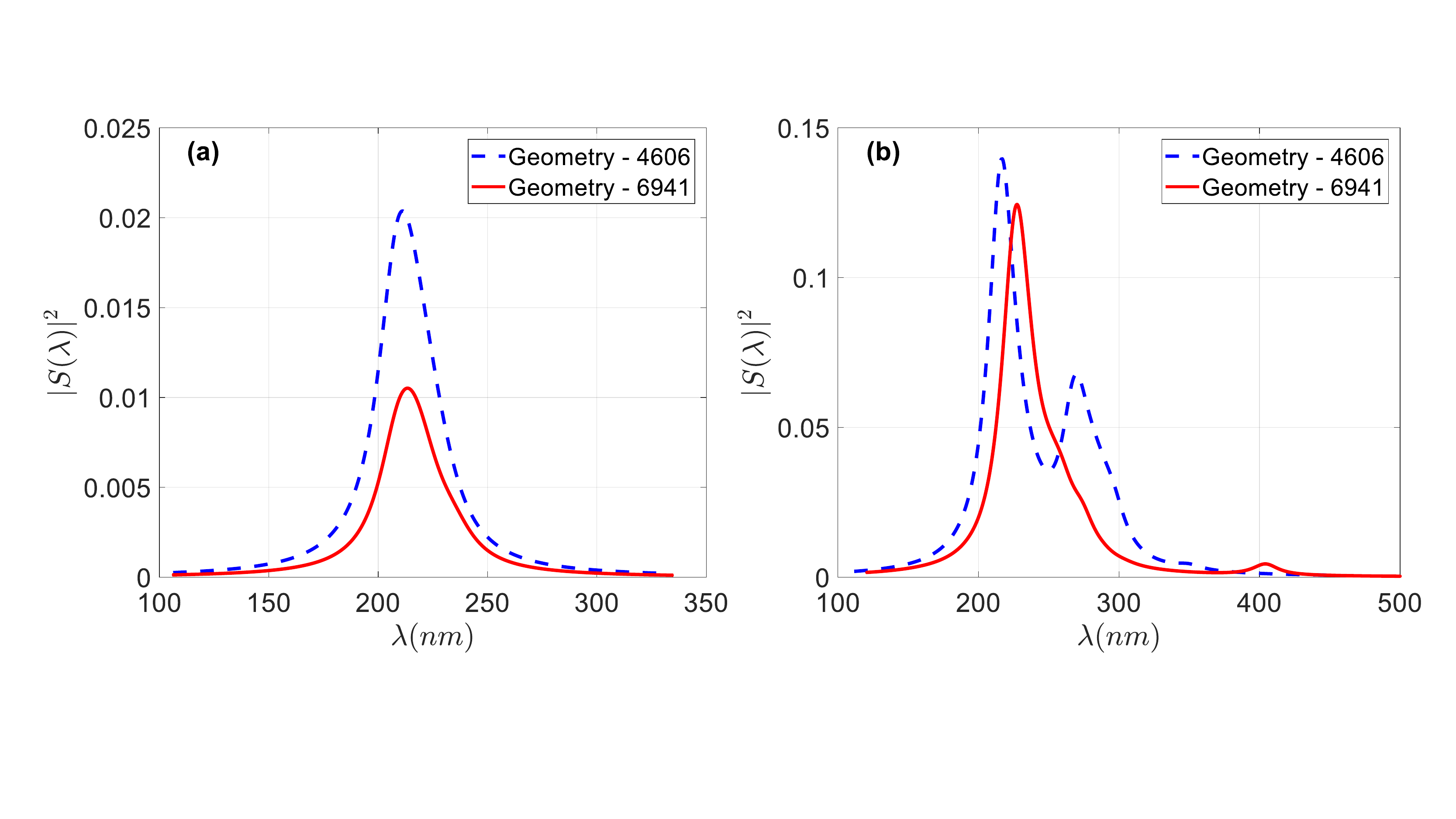}
	\caption{The absorption and emission Spectra of Cyclic-FF MD selected dimers, computed with
B3LYP-D3/6-311G**: (a) Absorption spectrum for geometries 4606 (dashed blue) and 6941 (solid red) - atoms are in their ground state minimum energy geometry ; (b) Emission spectra for geometries 4606 (dashed blue) and 6941 (solid red) - atoms are in the S$_1$ excited state minimum energy. Both geometries do not have a proton transfer in their excited state and, while their emission spectra differ from their absorption spectra, the difference is much smaller in comparison to the proton-transfer case.}
	\label{fig:MD_dimer_spectra}
\end{figure}

This reduction in de-excitation energy is a requirement for the fluorescence, but it does not predict
the emission intensities in the desired range. 
However, it is possible to argue that once the molecule has relaxed to the excited state geometry minimum, the only
requirement for the fluorescence to happen is that the electronic transition to the ground state is
allowed. The intensity of the emission is determined by the quantum mechanical transition
probability, which depends on the oscillator strength. The oscillator strength and the
corresponding absorption and emission spectra are plotted in Fig. 4. The absorption spectra are
computed with ground state optimized geometry while the emission spectra are computed using
the first excited state optimized geometry. We observe that the emission spectra shift considerably due to the
proton-transfer mechanism. The shift in the spectrum covers a significant part of the visible
region with considerable quantum mechanical transition amplitude. 
A detailed examination of the emission spectrum  reveals that the absorption
peak at $220nm$ has disappeared, while the next new peaks appear at 
$297nm$, $423nm$, $443nm$, and $532nm$.

We next examined two other dimer configurations, namely the MD4606 and MD6941 geometries, that were selected from the MD simulation. Those configurations are higher in energy
by around $2K_{B}T$ per monomer at room temperature. Neither configurations showed a proton transfer in the excited state minimization. The calculated absorption and emission spectra for
those structures are shown in Fig. 5. The absorption spectrum resembles that of the manual guess geometry, whereas the emission spectrum is significantly different. The MD4606 geometry retains the main absorption peak at
$220nm$ and shows a new peak at $270nm$ with a shallow shoulder between $300$ to $400 nm$. On the
other hand, the MD6941 geometry shows a small red-shift for the peak at 220nm and a much
weaker peak at 400nm. Overall, it is evident that the manual guess configuration, which
manifested a proton transfer in the excited state, had a significant gap reduction and a large
red-shift of the whole spectrum. In contrast, the other two, MD selected, configurations showed a
much less significant shift. This demonstrates that the proton-transfer happens only at certain
configurations of the Hydrogen bond between the monomers. A more thorough analysis is needed to 
estimate the yield of this effect at room temperature
or, equivalently, to estimate the percentage of dimers that are in a configuration that allows 
a proton transfer and fluorescence.

\subsection{Cyclic-FF Trimers}
In this section we examined larger aggregates of three molecules (i.e. trimers) of the Cyclic-FF peptide. We conducted MD simulations and selected the two lowest-energy configurations as initial configurations for the DFT and TDDFT calculations. Figure 6 shows the MD simulation energy versus time, as well as the location of the structures that were selected - MD2284 and MD4004. The selected structures were further relaxed to their energy minimum using ab-initio DFT B3LYP-D3/6-311G** calculations. The results for the binding energies of the resulting structures are given in Table 3. The binding energy of the lowest energy structure (i.e. structure MD4004) was found to be $0.024 Ha$ ($0.65eV$) per monomer, equivalent to $15.06 Kcal/mol$, which is slightly less than the lowest binding energy that was found for the dimer. The MD2284 structure binding energy was found to be $0.021 Ha$ ($0.57eV$) and about $3K_{B}T$ higher at room temperature.

Next, we calculated the excited state properties using the same procedure that we used for the dimer structures. The MD4004 geometry showed a proton transfer in the $S_{1}$ excited state and a significant change in the emission spectrum in comparison to the absorption spectrum - the latter has a single large peak at $220nm$, whereas the former shows stronger peaks at $300nm$ and $400nm$, as well as some weaker peaks at $470nm$, $560nm$ and $650nm$. The second geometry configuration, MD2284, does not show a proton transfer state and its calculated emission spectrum is only slightly shifted from the absorption spectrum, both showing a similar structure with a single peak at around $220nm$.

 \begin{figure}[htbp]
\centering
{\includegraphics[width=1.0\linewidth,trim={0cm 0cm 0 -1.5cm},clip]{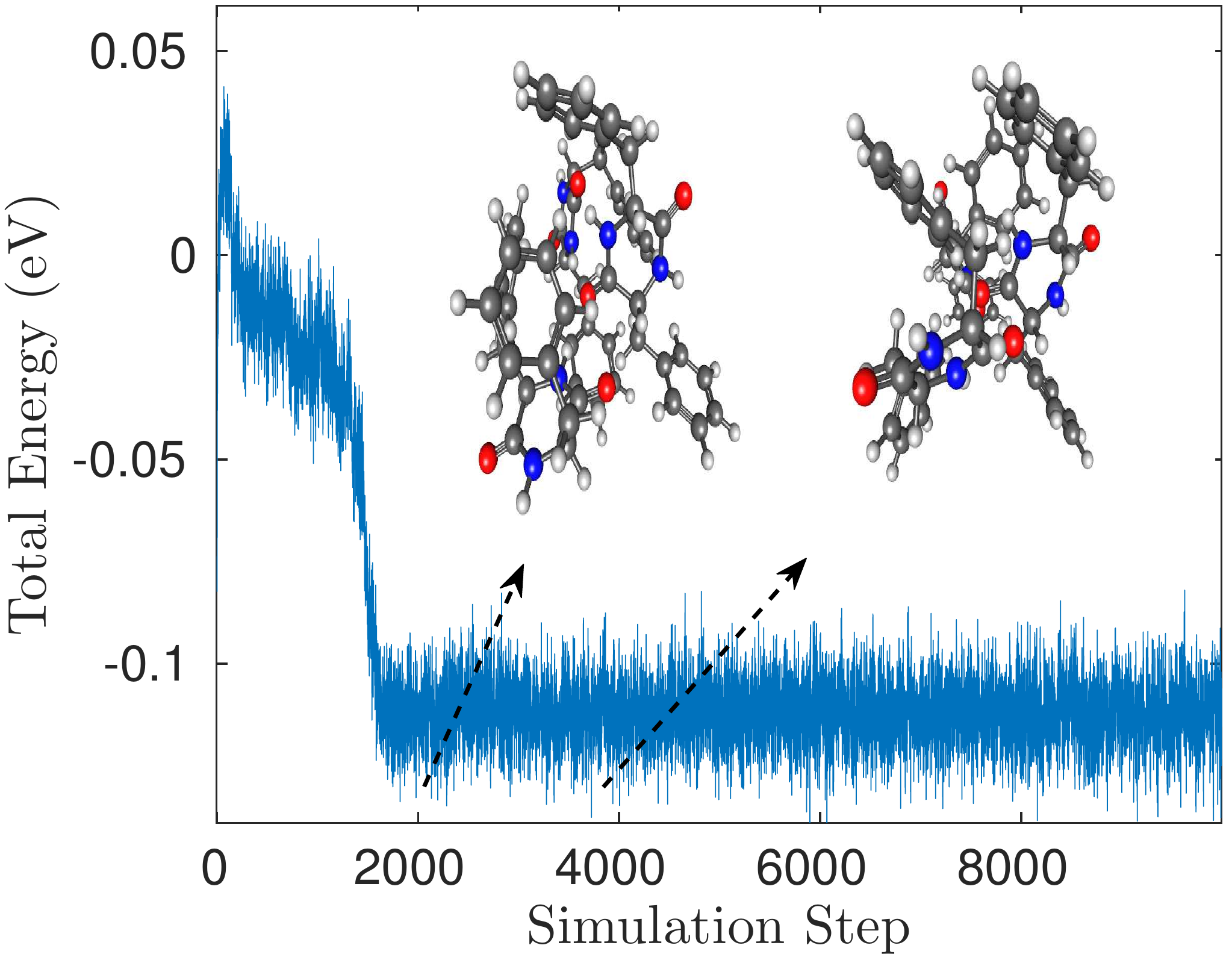}}
\caption{Cyclic-FF trimer MD geometries are shown. The different molecular
configurations' energy is shown as a function of the MD simulation step. Two of the lowest energy configurations were selected for further analysis and marked with black dotted arrows. The selected configurations are the 2284th and 4004th geometries, respectively.}
\label{fig:trimer_MD}
\end{figure}

\begin{table*}[!htbp]
	\begin{tabular}{ | l | l | l |} \hline
		Cyclic-FF Trimer & Cyclic-FF Trimer Ground State   & Binding Energy  \\ 
		Ground State Geometry &  Energy in Hartree ($Ha$)& per Molecule in Hartree ($Ha$)\\ \hline
		MD Geometry 2284   & -2871.1479  & -0.0210 \\ \hline
		MD Geometry 4004   & -2871.1569  & -0.0240  \\ \hline
		
	\end{tabular}
	\label{table:trimer_binding_energies}
	\caption{Trimer ground state total and binding energies}
\end{table*}

\begin{figure}[htbp]
 \centering
\includegraphics[width=1.0\linewidth,trim={1cm 3.5cm 0 2.5cm},clip]{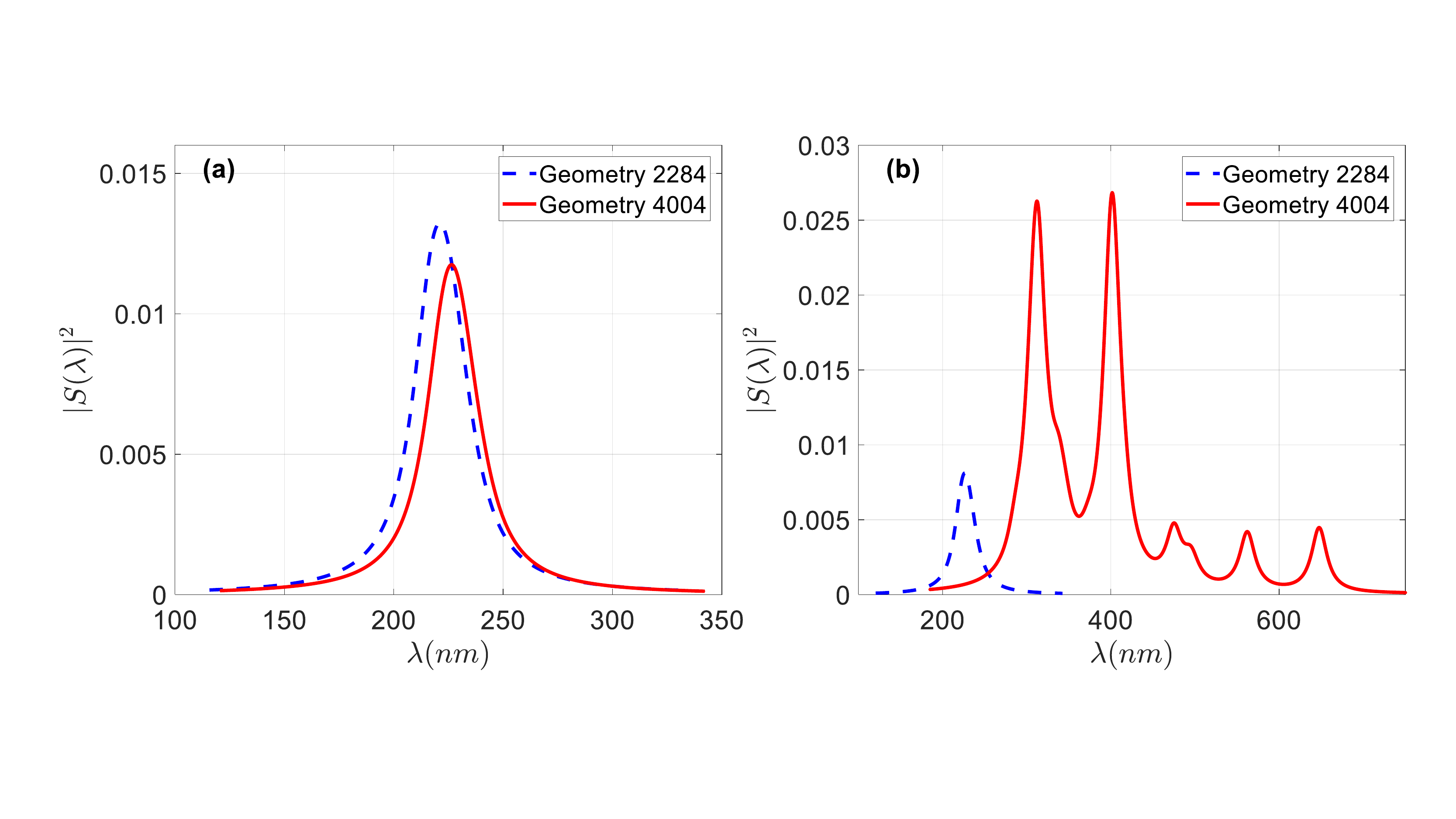}
 \caption{Absorption and emission spectra for Cyclic-FF trimers: (a) Absorption spectrum of geometries 2284 (dashed blue) and 4404 (solid red), atoms are in the ground state geometry energy minimum; (b) Emission spectrum of geometries 2284 (dashed blue) and 4404 (solid red), atoms are in the S$_1$ excited state geometry energy minimum. Geometry 4404, which shows a proton transfer in S$_1$, also exhibits a significant red-shift in the spectrum, Geometry 2284, which does not have a proton transfer, has an emission spectrum that is very close to that of the absorption.}
 \label{CFFSpectraMD}
 \end{figure}
\begin{table*}[!htbp]
\begin{tabular}{ | l | l | l |l|} \hline
MD Selected & Ground State &  Excited State & Energy Gap in $eV$ \\
Trimer Geometry & Energy ($E_{0}\, (H_{a})$) & Energy ($E_{n}\, (H_{a})$) & ($\Delta E_{n}=E_{n}-E_{0}$)\\ \hline
Ground State Optimized &  $E_{0}= -2871.1479$  & $E_{1}= -2870.9600$ & $\Delta E_{1}= 5.11\,$  \\ 
Trimer Geometry (2284) &   & $E_{2}=-2870.9574$ & $\Delta E_{2}=5.18\,$  \\ \hline
First Excited State           &             &             & \\
Optimization  & $E_{0}= -2871.1498$ & $E_{1}=-2870.9619$ & $\Delta E_{1}= 5.11\,$  \\
Trimer Geometry (2284) &             & $E_{2}=-2870.9593$ & $\Delta E_{2}= 5.18\,$  \\
(No-Proton Transfer)   &  &  & \\ \hline
Ground State   &  &  & \\
Optimization          &  $E_{0}=-2871.1569$ & $E_{1}=-2870.9631$ & $\Delta E_{1}=5.27 \,$\\ 
Trimer Geometry (4004)&  & $E_{2}=-2870.9626$  & $\Delta E_{2}= 5.29 \,$\\ \hline
First Excited State   &  & & \\
Optimization          &  &  & \\
Trimer Geometry (4004)& $E_{0}= -2871.0747$ & $E_{1}=-2871.0695 $ &  $\Delta E_{1}= 0.14 \,$  \\ 
(Inter-Molecular      &  & $E_{2}=-2871.0262$ &  $\Delta E_{2}= 1.32 \,$\\
Proton Transfer)      &  &  &  \\ \hline
\end{tabular}
\caption{Trimer ground and excited state energy gap (No proton-transfer \& proton-transfer)}
\label{table:Trimer_excitation_energies}
\end{table*}

\section{Summary}

We demonstrated that an excited state {\it inter}-molecular proton transfer can occur in dimers and trimers of cyclic FF molecules. This proton transfer takes place within the Hydrogen bonds between the monomers, and is the basis for a large red- shift in the emission spectrum. This red-shift change in the emission properties can explain the phenomenon of fluorescence in the visible range, which was recently observed in UV-light excitation measurements of Cyclic-FF molecule aggregates. The Hydrogen bonds are highly important for this phenomenon, due to two reasons. Firstly, these bonds stabilize the formation of dimers and trimers, and probably also for larger molecule aggregates~\cite{handelman2016reconstructive}. Secondly, Hydrogen bonds enable proton transfer without barrier crossing, along the Hydrogen bond, resulting in a modified geometry with lower energy in the excited state.
 
Such an effect of proton transfer induced fluorescence was also demonstrated in other molecular systems~\cite{RefESIPT_4,flur_ptran_htrans2017}. Our analysis also shows that the mechanism of proton transfer is highly dependent on the formation of specific Hydrogen bonds between the monomers.
Although we demonstrated this phenomenon with a di-peptide that possesses phenyl side groups, the mechanism is clearly dependent on the Hydrogen bonds and not on the aromatic nature of the side groups. As this mechanism is of a quite general nature we expect to find it also in aggregates of other peptides. Finally, we analyzed only the transitions that happen from $S_1$ and $S_2$ to $S_0$, excitation to higher states, $S_n$, could be accompanied also by a proton transfer and hence could lead to excitation dependent fluorescence.

\section{Acknowledgments}

We would like to thank the Israel Ministry of Science, Technology and Space for financial support.

\end{document}